# Anion emission from water molecules colliding with positive ions: Identification of binary and many-body processes

J.-Y. Chesnel,[1,*] Z. Juhász,[2] E. Lattouf,[1] J. A. Tanis,[1,3] B. A. Huber,[1] E. Bene,[2] S. T. S. Kovács,[2] P. Herczku,[2] A. Méry,[1] J.-C. Poully,[1] J. Rangama,[1] and B. Sulik[2]

[1]*Centre de Recherche sur les Ions, les Matériaux et la Photonique (CIMAP), Unité Mixte CEA-CNRS-EnsiCaen-Université de Caen Basse-Normandie, UMR 6252, 6 Boulevard Maréchal Juin, F-14050 Caen cedex 04, France.*

[2]*Institute for Nuclear Research, Hungarian Academy of Sciences (MTA Atomki), H-4001 Debrecen, P.O. Box 51. Hungary.*

[3]*Department of Physics, Western Michigan University, Kalamazoo, Michigan 49008 USA.*

*Corresponding author: jean-yves.chesnel@ensicaen.fr

It is shown that negative ions are ejected from gas-phase water molecules when bombarded with positive ions at keV energies typical of solar-wind velocities. This finding is relevant for studies of planetary and cometary atmospheres, as well as for radiolysis and radiobiology. Emission of both $H^-$ and heavier ($O^-$ and $OH^-$) anions, with a larger yield for $H^-$, was observed in 6.6-keV $^{16}O^+$ + $H_2O$ collisions. The experimental setup allowed separate identification of anions formed in collisions with many-body dynamics from those created in hard, binary collisions. Most of the anions are emitted with low kinetic energy due to many-body processes. Model calculations show that both nucleus-nucleus interactions and electronic excitations contribute to the observed large anion emission yield.

Anions are of fundamental interest in atomic physics since electron correlation effects are generally more important in negative ions than in atoms or positive ions. Moreover, anions play a significant role in various fields, ranging from astrophysics, atmospheric and plasma physics to surface physics and accelerator science [1-3]. Since anions influence appreciably the properties of the media in which they are present [1-9], knowledge of the mechanisms and routes leading to their formation is of prime importance.

Water is an essential and abundant molecule in a variety of planetary and space environments as well as in living media. Numerous studies have been devoted to anion emission from gas-phase water molecules following electron impact [10-15]. For other molecules, it has been shown that anion formation can be collisionally induced not only with electrons but also with positive ions as projectiles [16-19]. In this context, an important question is whether anions are ejected from free water molecules when bombarded by cations and, if so, to what extent. This is of interest for astrophysically-relevant collisions of stellar-wind ions [1-9] and for radiochemistry [20,21]. In dilute and dense media containing water, the subsequent interaction of negative fragments with neighboring molecules may create new species. For instance, the $H^-$ + $H_2O$ reaction may form $H_2$ and $OH^-$, and the $O^-$ + $H_2O$ reaction leads to OH and $OH^-$ [20]. Also, the addition of $H^-$ anions to DNA bases surrounded by $H_2O$ can result in proton migration, and subsequently to DNA damage [21].

More generally, the question arises as to which mechanisms are responsible for cation-induced anion formation. Previous collisional studies have shown specific cases of the formation of $H^-$ anions in soft collisions involving negligible momentum transfer between the collision partners [16-19]. There, the dissociation of the molecules was governed by electronic excitation and/or electron transfer. On the other hand, our recent studies [22,23] have shown that anions are also created in hard collisions involving an energetic encounter between two atomic cores (a "binary-encounter process") at impact energies of a few keV. Soft many-body collisions, in which interactions between all the atomic cores of the collision complex are of similar importance, are likely to contribute to anion formation as well. However, determination of the relative contribution of binary and many-body processes has not been done and remains challenging.

In this Rapid Communication, we show that many-body processes dominate the anion emission in few keV $O^+$ + $H_2O$ collisions giving a large yield of low-energy anions. The projectile energy corresponds to typical solar wind velocities (300 km/s). It is demonstrated that the contribution of anion emission is strongly enhanced by the interplay between nucleus-nucleus interactions and electronic excitation. Experimentally, contributions from the binary-encounter and the "soft-collision" processes were identified and measured separately. In binary collisions most of the emitted ions have high, angular-dependent kinetic energies, while soft many-body processes lead predominantly to lower emission energies in the laboratory (target) frame. The experimental set-up was developed to significantly reduce the otherwise dominant electron yield, in order to measure absolute *double-differential* cross sections (DDCS) for anion emission in the entire emission-energy range. On the theoretical side, classical model calculations were performed with a full description of the core-core interactions, including the kinetic energy released in electronic excitation and ionization processes. Good overall agreement is found between theory and experiment over the entire angular and energy range when both core-core interaction and electronic excitation were included in the calculations.

The experiment was performed at the Grand Accélérateur National d'Ions Lourds (GANIL) in Caen, France. A beam of 6.6 keV $^{16}O^+$ ions with a current of ~ 150 nA crossed an effusive gas jet of $H_2O$ with a density of ~$10^{11}$ cm$^{-3}$. The collision-induced ionic fragments were selected according to their kinetic energy per charge by means of a 45° parallel-plate electrostatic analyzer (Fig. 1) with an energy resolution of 5% and an angular acceptance of 2°. The spectrometer was mounted on a rotatable ring for setting the observation angle, $\theta$. The experimental setup was similar to the one depicted in Refs. [24,25] and additional relevant details on the experimental method and data analysis are provided in Ref. [23].

This set-up has been modified to avoid electron contamination of the anion spectra. The electrons and anions originating from the collision area were magnetically separated by deflecting electrons before entering the electrostatic analyzer and again after leaving it, so that they could not reach the detector (Fig. 1). At the entrance and the exit of the spectrometer (shaded areas in Fig. 1) magnetic fields of $10^{-4}$ T were sufficient to filter out electrons with kinetic energies below 2 keV. The magnetic fields were provided by pairs of coils which were designed to minimize the residual field inside the electrostatic analyzer.

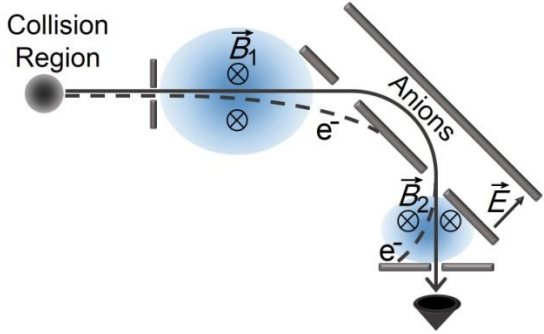

**Figure 1.** (Color online) Scheme of the electrostatic analyzer equipped with magnetic filters at its entrance and exit. Electrons are deflected so that only anions reach the detector.

The same apparatus was used for performing complementary time-of-flight (TOF) measurements, in order to determine the relative abundance of the different anionic species. To do so, the incident beam was pulsed at a period of 60 μs, with pulses of 2 μs in width. The mass-dependent time of flight of the particles from the collision area to the detector was measured when the spectrometer was set to a selected detection energy and angle. The flight distance was ~ 200 mm. These TOF measurements were also used to verify that the electron contribution at low emission energies was sufficiently reduced to be lower than the anion yield.

Figure 2 shows experimental DDCS with respect to the outgoing energy for various angles of anion emission in 6.6-keV $^{16}O^+$ + $H_2O$ collisions. These cross sections are double differential in emission energy and solid angle. The main component of each spectrum is a broad, slowly decreasing structure, which will be referred to as the *continuous spectrum*. At forward observation angles (< 90°) this component is superimposed by peaks at higher energies.

The continuous spectra maximize at low emission energies. There, the relative contributions of the different anion species were derived from TOF measurements. As the flight distance was ~ 200 mm, the pulse duration of 2 μs was short enough to separate the $H^-$ ions from the oxygen-containing anions ($O^-$ and $OH^-$) at emission energies ranging from 3 to 25 eV. To keep a satisfactory signal-to-noise ratio, no attempt was made to reduce the pulse duration (down to a fraction of μs) for a separate identification of the $O^-$ and $OH^-$ components. At forward and backward emission angles, the $H^-$ contribution is significantly larger than that of the heavy oxygen-containing anions, while in the 75°-90° angular range both contributions are about equal. The angular distribution of the light anions ($H^-$) is nearly isotropic, while that of the heavy anions is narrow, and centered near 90°. The latter behavior resembles binary collisions, where the slowest recoil ions move almost perpendicularly to the projectile trajectory. This is consistent with the picture that the collision of the two oxygen cores is close to a pure binary encounter, in which the light hydrogen cores play a minor role.

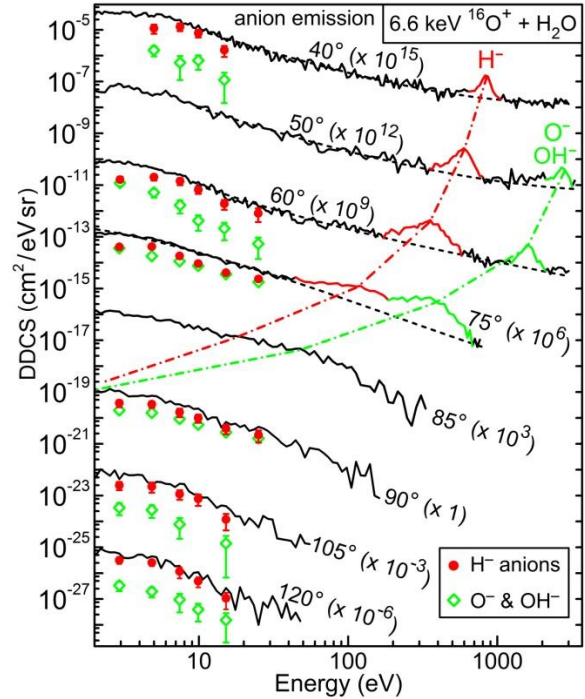

**Figure 2.** (Color online) Full curves: DDCS for anion emission in 6.6-keV $^{16}O^+$ + $H_2O$ collisions at the indicated observation angles. For graphical reasons, each spectrum is multiplied by the indicated factor. Anions formed in many-body collisions are observed in the slowly-decreasing "continuous" part of the spectra (full and dashed curves in black). Anions created in binary-encounter collisions appear in peak structures. The peak centroids are shifted towards lower energies as the angle is increased (graphically emphasized by the dash-dotted curves). The points at lower energies give the relative contributions due to $H^-$ (red circles) and $O^-/OH^-$ (green diamonds).

The role of many-body dynamics in low energy $H^-$ emission is expected to be significantly more important due to the larger perturbation induced by the heavy target oxygen atom on the trajectories of the light H fragments. This picture is supported by the fact that $H^-$ emission is not far from isotropic at low energies, thus resembling the emission of electrons in ionization [26].

Pronounced peaks at higher energy (Fig. 2) result from recoil anions. Their first observation goes back to the earlier study of $OH^+$ + acetone collisions [22]. They appear only at forward angles (< 90°). Their mean energy decreases strongly with increasing angle. From a simple kinematic calculation assuming an elastic two-body collision between the projectile and a single target atom, the peak at lower energy (in red) could be assigned as originating from $H^-$ and the other peak (in green) as to be due to $O^-$ and $OH^-$. These anions are formed in hard binary collisions occurring at small impact parameters with a large momentum transfer. As one of the receding target cores leaves the collision complex it may capture enough electrons to become an anion. The same applies for the projectile, and thus in a binary O-O collision the scattered projectile may also contribute to the $O^-$ peak.

In this experiment, DDCS spectra for cation emission were also measured (not shown) and, similar to the anion spectra, they exhibit recoil peaks at forward angles. From their areas, the single-differential cross sections (SDCS) for $H^-$ and $H^+$ emission via the binary process were determined. These cross sections are differential in emission solid angle. The angular dependence of both the $H^-$ and $H^+$ cross sections is compared with the theoretical cross section (see Ref. [22]) for the elastic recoil of an H atom by the impact of an O atom (Fig. 3a). The SDCS $H^-/H^+$ ratio is about constant, with an average value of $\sim 10^{-2}$ (Fig. 3b). The theoretical curves match the experimental data when multiplied by constant factors, which represent the relative populations of the different charge states. It is found that $(0.7 \pm 0.4)\%$ of the scattered H atoms becomes $H^-$ and $(60 \pm 33)\%$ of them becomes $H^+$.

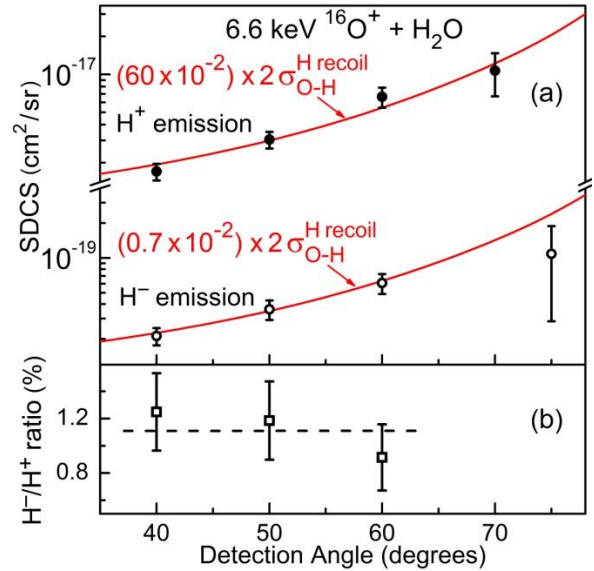

**Figure 3.** (Color online) **Part (a):** SDCS for $H^-$ (open circles) and $H^+$ (full circles) emission via the binary-encounter process as a function of the observation angle. Only *relative* error bars due to statistical uncertainties are shown, except the rightmost points, where a larger uncertainty stems from the overlap of peak structures. Red curves: calculated SDCS for two-body elastic recoil of H atoms by 6.6-keV oxygen impact, multiplied by factors representing the fraction of the different charge-state components. **Part (b):** Ratio between the experimental $H^-$ and $H^+$ SDCS reported in (a).

The results shown in Fig. 3 confirm our recent findings for $H^-$ emission via the binary process in $OH^+$ + Ar collisions at the same velocity [23]. Namely, the relative populations of the different charge states of the hydrogen fragments do not depend significantly on the emission angle, the impact parameter or the momentum transferred between the collision partners. Hence, we have found that the charge state distribution of the hydrogen fragments resulting from the binary process is akin to a statistical distribution in different molecular collisions [22,23]. This statistical aspect likely stems from the fact that the number of possible final states for the outgoing fragment is very limited compared to the number of electronic transitions which may occur in each collision.

By integrating the double differential spectra over energy and solid angle (Fig. 2), the total cross section for anion formation in 6.6-keV $^{16}O^{+}$ + $H_2O$ collisions was found to be $(5 \pm 3) \times 10^{-18}$ cm$^2$. From the TOF measurements and from the binary peaks, we estimate that about two thirds of the detected anions are $H^-$ ions. Hence, the total cross section for producing $H^-$ ions in the present collision is estimated to be ~ $3 \times 10^{-18}$ cm$^2$, while it is ~ $2 \times 10^{-18}$ cm$^2$ for oxygen-containing anions. Here, it is of prime interest to compare the present cross sections for cation-induced anion formation with those for producing anions from water molecules in interaction with free electrons. Total cross sections for dissociative electron attachment (DEA) to water in the gas phase were previously measured as a function of the kinetic energy of the incident electrons [14]. In these measurements, resonances appeared as three peaks in the electron energy range from 5 to 15 eV. For $H^-$ formation by DEA to water, these peaks show maximum cross sections of about $5 \times 10^{-18}$ cm$^2$, $1 \times 10^{-18}$ cm$^2$ and $0.1 \times 10^{-18}$ cm$^2$ at electron energies of 6.5, 8.6 and 11.8 eV, respectively. Cross sections for $O^-$ formation by DEA to $H_2O$ do not exceed $0.3 \times 10^{-18}$ cm$^2$ in the electron energy range from 4 to 20 eV [14]. Hence, comparison of these previous data with the present ones shows that cross sections for cation-induced anion emission from $H_2O$ can be as large as those for anion formation by slow-electron impact. This finding may have significant implications in radiochemistry since ion-induced damage in matter stems from both primary ions and secondary electrons and radicals. Our data show that anions resulting from the impact of both primary and secondary cations on $H_2O$ can be formed in non-negligible yields compared to those of anions created by attachment of slow secondary electrons. Thus, cation-induced anion formation has to be taken into account for accurate modeling of reaction networks in a variety of water-containing media, such as atmospheres and living cells.

In our previous work on anion emission from acetone [22], only the anions formed via the hard binary-encounter process were observed. It was expected that emission in softer collisions may be even more significant. In the present work this expectation is confirmed by measuring the anion-production DDCS in the *entire* emission-energy range. By integrating the continuous part of the spectra over energy and angle, we show that anion formation via many-body processes accounts for about 90% of the total cross section.

The development of the theory of anion formation in cation-molecule collisions is hampered by the multi-electronic character of the process. Instead, numerical simulations of the trajectories of the different fragments are presently performed in order to interpret the experimental data. The method is similar to the one used in [23]. A two-body interaction between each pair of atoms is assumed. For each pair, the interaction potential is determined as a function of the internuclear distance by performing an ab initio calculation using the MOLPRO code [22,27]. Each potential refers to the relaxed ground-state energy of the diatomic systems. The trajectories of all the atomic cores are calculated by solving the coupled Newton equations of motion. Random initial conditions for the position of the projectile and for the orientation of the target molecule are used. By repeating this calculation for a large number of collisions (~$3 \times 10^6$), the energy and angular distributions of the ejected fragments are determined.

The sole goal of the model was to describe the dynamics of the atomic cores. Accordingly, no information was provided about the final charge states of the fragments. Since the anion and cation signals are not far from being proportional in the entire energy range, cross sections for emission in a particular charge state were derived approximately by multiplying the obtained cross sections by an appropriate constant factor. So, to calculate the cross sections for H⁻ emission, the obtained hydrogen emission cross sections were multiplied by the factor of $0.7 \times 10^{-2}$, *i.e.,* the H⁻ fraction derived from Fig. 3. In this way, it is ensured that the intensity of the calculated binary peak matches the experiment.

DDCS for H⁻ emission have been calculated. As shown in Fig. 4 (dotted curves), the high-energy peak is narrower in the calculation. Moreover, the cross sections are strongly underestimated at lower emission energies. This latter disagreement originates from the fact that less than 10% of the OH bonds are broken in the simulated collisions occurring at impact parameters larger than ~ 1 atomic unit. In such soft collisions, the kinetic energy transferred to H is lower than the dissociation energy of an OH bond of $H_2O$ in its ground state.

The observed high dissociation yields may however be described by introducing kinetic energy release (KER) into the model. This KER most likely stems from electronic processes such as dissociative excitation and ionization. For energetic collisions involving many centers, the number of dissociative channels can be extremely large. Hence, the KER should have a (rather wide) statistical distribution to describe the multitude of different excitation processes. Previous measurements on the collision-induced dissociation of $H_2O$ molecules show that the kinetic energy released in the cleavage of an OH bond is on the order of 5 eV and can even exceed 20 eV when highly excited states are involved [28]. Here, to reasonably match these previous data, the KER ($\approx \frac{1}{2} m_H v_{KER}^2$) is assumed to be a random variable with a Gaussian distribution centered at 5 eV with a standard deviation of 4 eV. Under these conditions, as negative KER values were excluded, the KER value ranged from 0 to 10 eV in 88% of the simulated collisions, while KER values larger than 15 eV occurred in less than 0.7% of the simulated events. As in [23], KER is introduced by adding a velocity component along the OH axis, $\vec{v}_{KER}$, to the velocity of the H atom when the distance between the projectile and the active H atom is minimum.

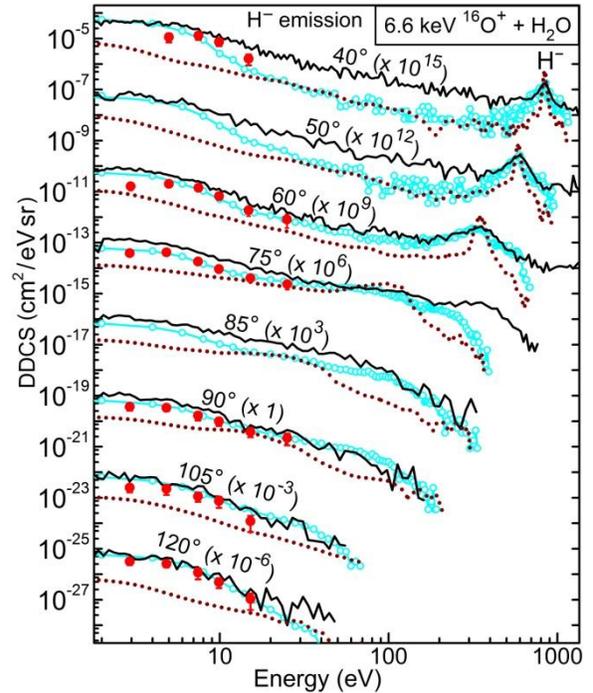

**Figure 4.** (Color online) **Open circles:** Simulated energy distribution of the H⁻ anions at different angles with KER; **Dotted curve:** Same without KER; **Full curve:** Experimental DDCS for anion emission; **Red, full circles:** Experimental H⁻ contribution from TOF measurements. Each spectrum is multiplied by the indicated factor.

In Fig. 4 the calculated cross sections for H⁻ emission (open circles) are compared with the experimental data. At low emission energy the introduced KER strongly enhances the simulated cross sections, so that the calculated curves agree fairly well with the data points obtained from TOF measurements. In the high energy part, the calculated recoil $H^-$ peaks reproduce satisfactorily the experimental ones in both width and amplitude. Though only a qualitative description was expected from this simulation, a satisfactory overall agreement was obtained. The crucial step towards this agreement is the introduction of an approximate KER distribution. According to the model, without kinetic energy release due to electronic excitation or ionization, the total cross section for $H^-$ emission would be about 3 times smaller.

In summary, an unexpectedly high yield of anion production, dominated by $H^-$ emission, was found in $O^+ + H_2O$ collisions at a typical solar-wind velocity. Double-differential cross-section measurements allowed separate identification of the two-body and multi-body processes and determination of their yields in the entire emission energy range. These DDCS measurements show that anion emission is concentrated at low emission energies. Model calculations including kinetic energy release show that electronic excitation and ionization processes play a decisive role in $H^-$ formation. The present absolute cross sections will aid in the understanding of the reaction pathways in various media in which cations interact with water. These findings have significant implications in collision physics and chemistry since slow anions are efficient agents for charge transfer and chemical reactions. The present results are also of importance in view of new emerging technologies with cryogenic anion storage rings, which will allow studies of inherent anion properties and their reactions [29].

Assistance from S. Guillous, A. Leprévost, F. Noury, J.-M. Ramillon and Dr. P. Rousseau is gratefully acknowledged. This work was supported by the Transnational Access ITS-LEIF, granted by the European Project HPRI-CT-2005-026015, the Hungarian National Science Foundation OTKA (Grant No. K109440), the French-Hungarian Cooperation Program PHC Balaton (Grant No. 27860ZL/TéT_11-2-2012-0028), and the French-Hungarian CNRS-MTA Cooperation (ANION project, Grant No. 26118). J.A.T. was supported by a grant from the Basse-Normandie region (Convention No. 11P01476) and from the European Regional Development Fund (Grant No. 32594).